\def\draftversion{false}
\RequirePackage{ifthen}
\ifthenelse{\equal{\draftversion}{true}}{
  \documentclass[aps,showpacs,prl,galley,superscriptaddress,
                 longbibliography]{revtex4-1}
}{
  \documentclass[aps,showpacs,prl,reprint,superscriptaddress,
                 longbibliography]{revtex4-1}
}

\usepackage[colorlinks=true,linkcolor=blue,citecolor=blue,urlcolor=blue]{hyperref}

\usepackage{hyperref}
\usepackage{setspace} %to set 1.5 spacing in text
\usepackage{graphicx}
\usepackage{amsmath}
\usepackage[usenames]{color}
\usepackage{amsmath}
\usepackage{amssymb}
\usepackage{verbatim}
\usepackage{latexsym}
\usepackage{enumerate} % to change the numbering of 'enumerate'
\usepackage{bm} % for bold face mathematical symbols

\usepackage{soul}  % \st    for strike-out

\ifthenelse{\equal{\draftversion}{true}}{
  \marginparwidth 2.7in
  \marginparsep 0.5in
  \newcounter{comm} % counter for commentaries
  \def\commnext{\stepcounter{comm}}
  \def\commtext{{\bf\color{blue}[\arabic{comm}]}}
  \def\commmar{{\bf\color{blue}[\arabic{comm}]}}
  \def\dvm#1{\commnext\marginpar{\small DV\commmar: #1}\commtext}
  \def\tmm#1{\commnext\marginpar{\small BM\commmar: #1}\commtext}
  \def\mlab#1{\marginpar{\small\bf #1}}
  
}{
  \def\dvm#1{}
  \def\tmm#1{}
  \def\mlab#1{}
  
}
\begin{document}

\title{Temperature effects in the band structure of topological insulators}
%\title{Temperature effects in the topological band structure of
%    the Bi$_2$Se$_3$ family of compounds}
%\title{Temperature-induced band inversion in topological insulators} 

\author{Bartomeu Monserrat}
\email{monserrat@physics.rutgers.edu}
\affiliation{Department of Physics and Astronomy, Rutgers University,
  Piscataway, New Jersey 08854-8019, USA}
\affiliation{TCM Group, Cavendish Laboratory, University of Cambridge,
  J.\ J.\ Thomson Avenue, Cambridge CB3 0HE, United Kingdom}
\author{David Vanderbilt}
\affiliation{Department of Physics and Astronomy, Rutgers University,
  Piscataway, New Jersey 08854-8019, USA}

\date{\today}

\begin{abstract}
We study the effects of temperature on the band structure
of the Bi$_2$Se$_3$ family of topological insulators using
first-principles methods. Increasing temperature drives these
materials towards the normal state, with similar contributions from
thermal expansion and from electron-phonon coupling. The band gap
changes with temperature reach $0.3$~eV at $600$~K, of similar
size to the changes caused by electron correlation. Our results
suggest that temperature-induced topological phase transitions should
be observable near critical points of other external parameters.
\end{abstract}

\pacs{}

\maketitle

% Introduction
%\section{Introduction}

Topological insulators and related materials have consolidated as a new
field of condensed matter of both fundamental and applied interest.
Potentially novel applications have motivated the search for new
materials exhibiting topological properties. In this quest,
composition~\cite{bisbx_ti_hasan},
pressure~\cite{pressure_ti_bitei,sb2se3_pressure_exp},
strain~\cite{strain_bi2se3,rappe_feti}, and electromagnetic
fields~\cite{efield_sb2te3_theory,efield_sb2te3_exp,efield_ti_zunger}
have all been used to control topological order.
%
%and first-principles calculations have proved central in
% predicting topological materials that have been subsequently
% verified experimentally.
%
By contrast, temperature has so far exclusively played a passive role, for
example constraining the quantum anomalous Hall effect to the milikelvin
regime~\cite{qah_experiment}, and thus limiting potential applications
of the associated dissipationless currents. Only recently have Saha
and co-workers proposed that temperature can also be exploited when
studying topological materials, for example by using phonon linewidths 
to identify band inversions~\cite{garate_ph_linewidths_band_inversion}.

The effects of temperature on topological insulators
can be divided into two areas. The first concerns the electronic
occupation of states, leading to questions such as the proper definition
of topological invariants, which must then be based on the density
matrix rather than the ground state wave function~\cite{uhlmann_phase_1d,
uhlmann_phase_2d_huang,uhlmann_phase_2d_viyuela,topo_density_matrices_diehl}.
The second relates to the renormalization of the single-particle bands, which could 
lead to temperature-induced band inversions~\cite{elph_topological_prl,
elph_topological_prb}, and it has been argued on the basis of simple models 
that temperature should favor topological phases.

First-principles calculations have proved central in predicting new
topological materials and their behaviour, many times leading to
subsequent experimental observation~\cite{bi2se3_nat_phys,type2_weyl}.
The study of temperature effects on topological materials would 
equally benefit from first-principles calculations, but none have 
been available.
Fully first-principles calculations of the
temperature dependence of band structures have only recently become
possible, as first and second order terms
in the electron-phonon interaction contribute similarly and must
be included on an equal footing~\cite{0022-3719-9-12-013,
giustino_elph_review_arxiv}.
Because of such difficulties, to date these calculations have only
been performed on simple semiconductors and 
insulators~\cite{PhysRevLett.105.265501,PhysRevLett.107.255501,
giustino_nat_comm,monserrat_elph_diamond_silicon,gonze_marini_elph}.
To use these techniques for topological materials, they need to be
extended to include the effects of spin-orbit coupling in the
calculation of the electron-phonon interaction.

In this work, we study the effects of temperature on topological 
insulators with first-principles methods. Using the
Bi$_2$Se$_3$ family of topological insulators, we demonstrate the 
feasibility of such calculations and unravel the importance of 
temperature when studying topological order. Our calculations show
that both thermal expansion and electron-phonon coupling make 
similar contributions to the temperature dependence of band 
structures, and that, at variance with predictions from model
studies~\cite{elph_topological_prl,elph_topological_prb}, 
increased temperature tends to suppress topological order in
these materials.

% Methods
%\section{Methods}

We consider the Bi$_2$Se$_3$ family of compounds for our 
first-principles study~\cite{bi2se3_nat_phys}.
Bi$_2$Te$_3$, Bi$_2$Se$_3$, and Sb$_2$Te$_3$ crystallize in the
rhombohedral $R\overline{3}m$ space group, with five atoms in the
primitive cell. They form a layered structure, in which groups of five
layers are tightly bound (so-called quintuple layers), and these
are then weakly bound to each other. Sb$_2$Se$_3$ has $Pnma$
symmetry in its ground state. Metastable structures of these compounds
have been synthesized~\cite{bi2se3_polymorphs}, and in particular the
$R\overline{3}m$ structure of Sb$_2$Se$_3$ has been experimentally
studied~\cite{sb2se3_pressure_exp}. In this work, we consider the
$R\overline{3}m$ structure for all compounds.

Our calculations are based on density functional theory (DFT) using the
{\sc vasp} package~\cite{vasp1,vasp2,vasp3,vasp4}. We use the PBE
functional~\cite{PhysRevLett.77.3865} and the projector augmented-wave
method~\cite{paw_original,paw_us_relation} with an energy cut-off of
$600$~eV and a $\mathbf{k}$-point grid of size $12\times12\times12$ for
the primitive cells and commensurate grids for the supercells. All
calculations are performed with spin-orbit coupling unless 
otherwise stated. 

The PBE functional overestimates static volumes and thermal expansion
compared to experiment. For example, in Bi$_2$Se$_3$ the experimental
hexagonal $c$ axis at $10$~K is
$28.48$~\AA\@~\cite{th_expansion_bi2se3}, compared with the PBE $c$ axis
of $30.21$~\AA\@. Therefore, we use experimental volumes when
available. For Bi$_2$Se$_3$ and Sb$_2$Te$_3$, experimental data is only
available up to about $250$~K~\cite{th_expansion_bi2se3}, and we fit
a Bose factor to this data to extrapolate to higher temperatures,
which is justified in this case as the linear asymptotic regime is
already reached at the highest temperatures for which data is
available~\cite{analytic_prb}. For Bi$_2$Te$_3$, data is available up to
$600$~K~\cite{th_expansion_bi2te3}. No data is available for the
rhombohedral structure of Sb$_2$Se$_3$, so in this case we use the
structure relaxed using the PBE functional. The atomic coordinates of
all atoms are relaxed until the forces are smaller than
$10^{-3}$~eV/\AA\@.

The phonon and electron-phonon coupling calculations are performed
using the finite displacement method in conjunction with the recently
developed nondiagonal supercell approach~\cite{non_diagonal}. Finite 
displacement methods allow us to straight-forwardly incorporate the 
effects of the spin-orbit interaction on electron-phonon coupling which, 
as far as we are aware, has never been attempted in this context. 
The reported results correspond to a 
vibrational Brillouin zone sampling using grid sizes of $4\times4\times4$, 
as calculations for Bi$_2$Se$_3$ using grids of size $8\times8\times8$ 
show that the electron-phonon induced band gap corrections are converged 
to better than $8$~meV at all temperatures considered.
The matrix of force constants is constructed by considering small
positive and negative symmetry-inequivalent atomic displacements of
amplitude $0.005$~\AA\@. The dynamical matrix,
obtained by a Fourier transformation, is diagonalized to calculate
the vibrational frequencies $\omega_{\mathbf{q}\nu}$ %and eigenvectors
%$\mathbf{e}_{\mathbf{q}\nu}$ 
at each phonon wave vector $\mathbf{q}$ and
branch $\nu$. %These are used to construct the normal modes of
%vibration, of amplitude $u_{\mathbf{q}\nu}$, along which the band gap at
%temperature $T$ is calculated as
The band gap at temperature $T$ is then calculated as
\cite{0022-3719-9-12-013,PhysRevLett.105.265501,PhysRevLett.107.255501,
giustino_nat_comm,monserrat_elph_diamond_silicon,gonze_marini_elph,
giustino_elph_review_arxiv}
\begin{equation}
E_{\mathrm{g}}(T)=
E_{\mathrm{g}}^{\mathrm{static}}+
\frac{1}{N_{\mathbf{q}}}\sum_{\mathbf{q},\nu}\frac{a_{\mathbf{q}\nu}}{\omega_{\mathbf{q}\nu}}
\left[\frac{1}{2}+n_{\mathrm{B}}(\omega_{\mathbf{q}\nu},T)\right],
\end{equation}
where $a_{\mathbf{q}\nu}$ are the electron-phonon matrix elements averaged
over $N_{\mathbf{q}}$ $\mathbf{q}$-points, and
$n_{\mathrm{B}}$ is a Bose-Einstein factor. The electron-phonon matrix
elements are determined using finite displacements along the normal
modes of vibration.

% Results
%\section{Results}

\begin{figure}
\includegraphics[scale=0.45]{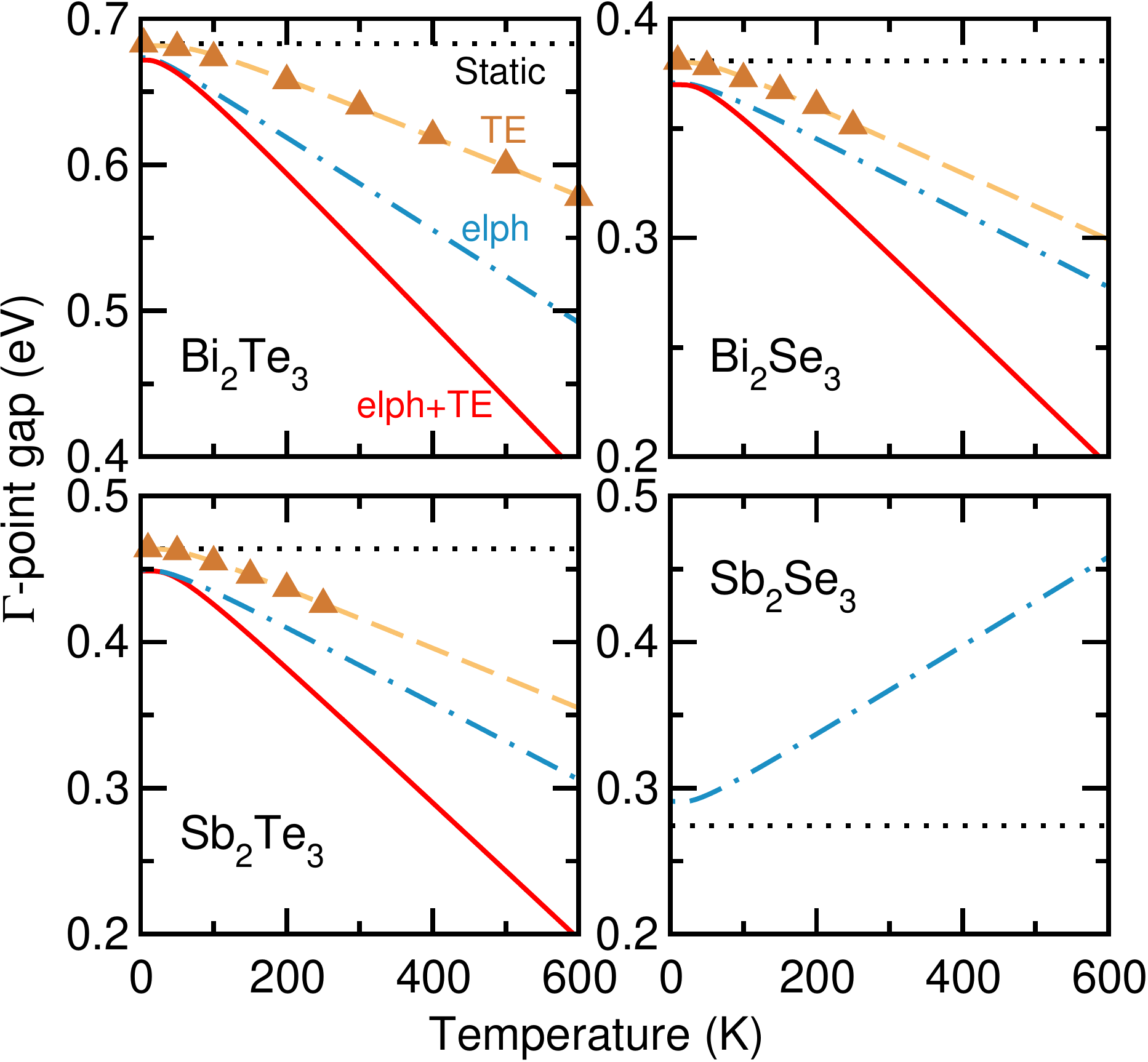}
\caption{Temperature dependence of the $\Gamma$-point band gaps
of the Bi$_2$Se$_3$ family of compounds. The black dotted lines
correspond to the static lattice band gaps; the orange dashed and
blue dashed-dotted lines show the result of including only
thermal expansion or only electron-phonon coupling effects,
respectively; and the red solid curves give the total 
temperature dependence.}
\label{fig:tdep}
\end{figure}

The topological nature of the Bi$_2$Se$_3$ family of compounds is
determined by the parity of the occupied bands at the $\Gamma$ point,
where band inversion occurs in the presence of spin-orbit
coupling~\cite{bi2se3_nat_phys}. Therefore, in order to investigate the
effects of temperature on the topological order, we focus on the
temperature dependence of the band gap at $\Gamma$, as shown in
Fig.~\ref{fig:tdep}.

Bi$_2$Te$_3$, Bi$_2$Se$_3$, and Sb$_2$Te$_3$ are
topological insulators, and their gaps close with increasing
temperature. Using Bi$_2$Se$_3$ as an example, we observe that 
the gap closure is driven by both thermal expansion and
electron-phonon coupling. Thermal expansion alone (orange dashed
line) closes the band gap by almost $0.1$~eV at $600$~K. 
Electron-phonon coupling alone (blue dashed-dotted line) 
causes a change in the band gap of $0.1$~eV at
$600$~K. The asymptotic linear dependence of the gap change with 
temperature is reached at around $100$~K, determined by the 
low-energy vibrations that are a consequence of the heavy atoms 
in the compounds under study. 
Thermal expansion and electron-phonon coupling make similar 
contributions, and therefore both must be included (red solid line). 
To date, the vast majority of first-principles calculations of
the temperature dependence of the band gap have neglected the
effects of thermal expansion.
This may be justifiable for simple materials made of light atoms,
such as diamond and silicon, that exhibit weak thermal expansion.
By contrast, topological 
insulators are composed of heavy elements, for which thermal
expansion is important as shown in Fig.~\ref{fig:tdep}.

Not surprisingly, for Sb$_2$Se$_3$, which is already a normal
insulator with an uninverted gap at low temperature, the same
electron-phonon effects lead instead to a band gap opening with
increasing temperature.
Thermal expansion also tends to increase the band gap,
but we have not included this contribution as there is no experimental
data available for this system.

The sizes of the band gap shifts are similar to those observed and 
calculated in many semiconductors and insulators. The only reported
phonon-induced gap
shifts in topological insulators we are aware of are those of Kim and
Jhi, who theoretically studied
the change in the band gap induced by exciting selected
phonon modes in a series of IV-VI semiconductors and found shifts on the
$0.1$~eV scale~\cite{elph_topological_jhi}. In the Bi$_2$Se$_3$ family
of compounds, zero-point contributions are small, in the range
$0.01$-$0.02$~eV in all systems, but thermal changes are larger,
reaching $0.2$-$0.3$~eV at $600$~K. The $GW$ approximation has been
shown to correct the DFT gaps by about $0.2$-$0.3$~eV in the
Bi$_2$Se$_3$ family of compounds~\cite{gw_bi2se3_louie}, to lead to
qualitative changes in the shape of the bands near
$\Gamma$~\cite{gw_bi2se3_blugel}, and more generally to question the
validity of some predictions of topological insulators based on
semilocal DFT~\cite{zunger_false_positives}. Our results show that the
effects of temperature
%(both thermal expansion and electron-phonon coupling) 
modify the band structure to a similar extent, and therefore
should also be included for an accurate description of topological
insulators.

\begin{figure}
\includegraphics[scale=0.45]{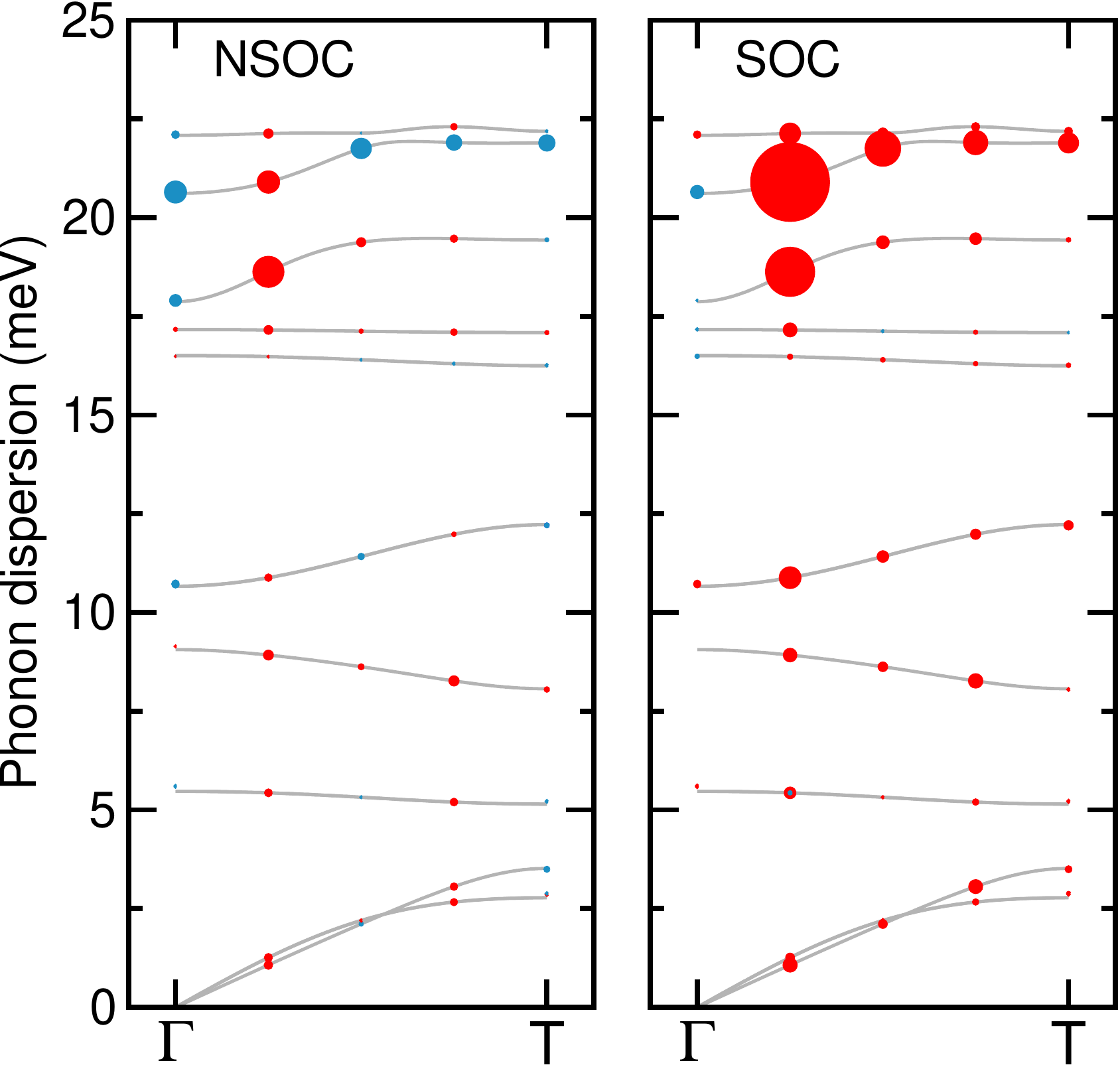}
\caption{Phonon dispersion of Bi$_2$Se$_3$ with mode resolved
electron-phonon coupling strength without (left) and with (right)
spin-orbit coupling. The area of the circles are proportional to the
strength of electron-phonon coupling, and the sign of the coupling is 
positive (gap opening) for blue circles and negative (gap closing) 
for red circles.} \label{fig:elph}
\end{figure}

We next investigate the microscopic origin of the electron-phonon
coupling strength, taking Bi$_2$Se$_3$ as an example. In
Fig.~\ref{fig:elph}
we show the phonon dispersion of Bi$_2$Se$_3$ along
the high-symmetry line from $\Gamma$ at $\mathbf{q}=(0,0,0)$ to
$\mathrm{T}$ at $\mathbf{q}=(0.5,0.5,0.5)$, chosen because
electron-phonon coupling is strongest along this line. The mode-resolved
strength of electron-phonon coupling to the direct band gap at the
electronic $\Gamma$ point is indicated by the filled
circles, and shown both without and with spin-orbit coupling. The
inclusion of the spin-orbit interaction has two effects. The first is to
reverse the gap-opening induced by electron-phonon coupling to gap
closing upon band inversion. The second is to modify the strength of
electron-phonon coupling beyond the mere exchange of band extrema. For
example, the gap closing induced by the modes at
$\mathbf{q}=(0.25,0.25,0.25)$ with spin-orbit coupling is larger than
the opening of the gap induced by the same modes when spin-orbit
coupling is not included. For Bi$_2$Te$_3$ and Sb$_2$Te$_3$, similar
behaviour is observed, but for Sb$_2$Se$_3$, the inclusion of spin-orbit
coupling does not induce a band inversion, and therefore the sign of the
band gap correction does not reverse.

A detailed analysis of the couplings reveals that the modes that couple
most strongly to the electronic states are those that involve atomic
vibrations along the $c$-axis that change the interlayer distance. It
is also the change in length of the $c$-axis that drives the band gap
change due to thermal expansion. This can be explained by the nature of
the states at the band extrema of the $\Gamma$ point, which mainly arise
from the hybridized $p$ orbitals of the constituent atoms. The crystal
field splits the $p_{z}$ from the $p_{x,y}$ orbitals, which remain
degenerate, and it is the $p_{z}$ that form the band extrema and undergo
band inversion when the spin-orbit interaction is
included~\cite{bi2se3_nat_phys}. Changing the interlayer distance,
either by atomic vibrations or by lattice expansion, modifies the states
at the gap edges, and thus drives the observed temperature dependence.

\begin{figure}
\includegraphics[scale=0.45]{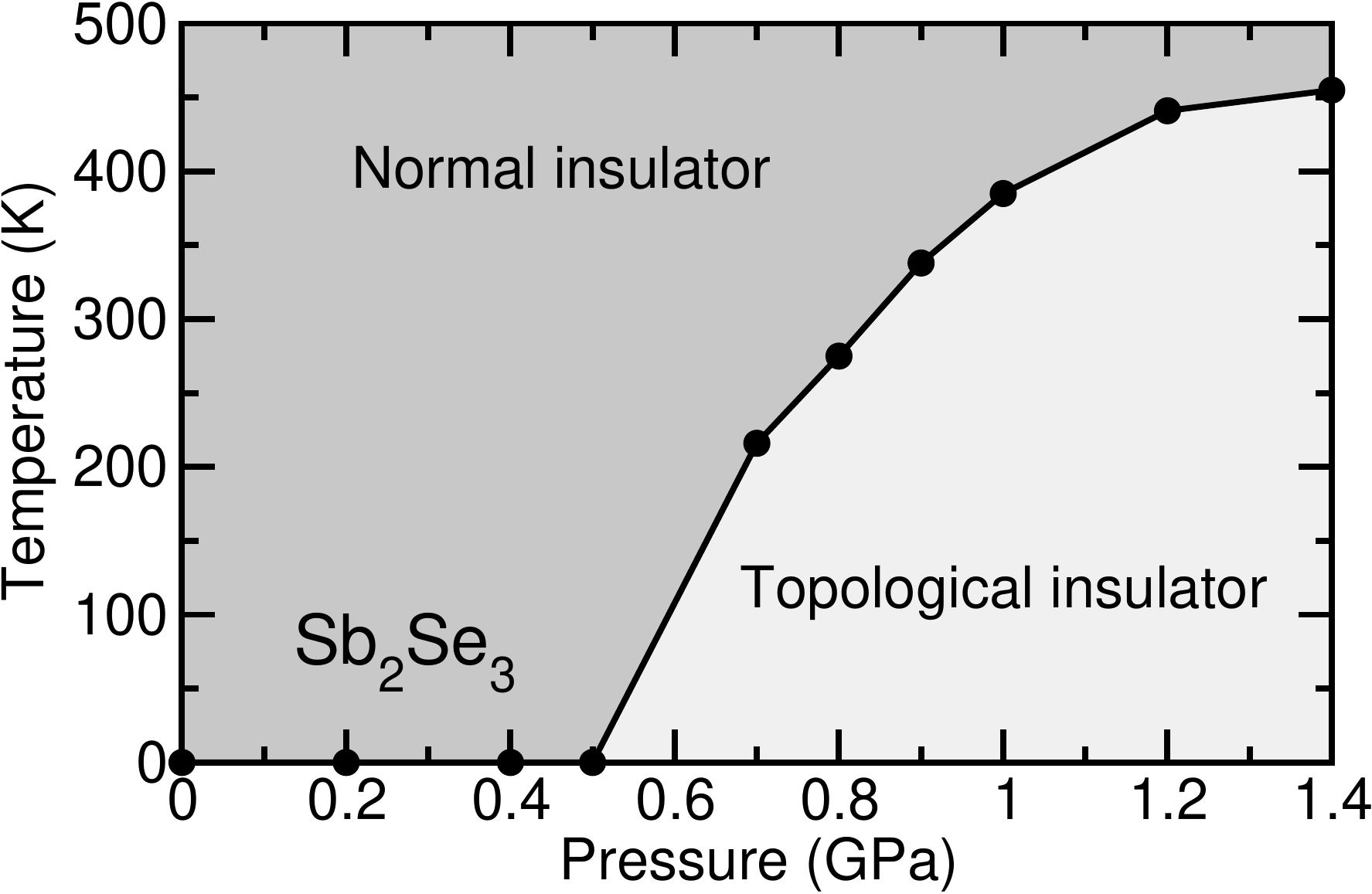}
\caption{Pressure-temperature phase diagram of Sb$_2$Se$_3$.}
\label{fig:phasediagram}
\end{figure}

The inclusion of temperature in our calculations allows us to construct
the pressure-temperature phase diagram of Sb$_2$Se$_3$ shown in
Fig.~\ref{fig:phasediagram}.  Sb$_2$Se$_3$ is a normal insulator under
ambient pressure, but with increasing external pressure it undergoes a
phase transition to a topological insulator at
$2.5$~GPa~\cite{sb2se3_pressure_exp}. First-principles calculations
based on the PBE functional give a transition pressure of
$1.0$~GPa~\cite{sb2se3_pressure_theory}. Our own calculations using PBE
give a similar transition pressure of $0.5$~GPa, and we also
perform HSE calculations~\cite{hse06_functional,hse06_functional_erratum} 
which increase the transition pressure to $2.5$~GPa, in better agreement with
experiment. Due to the computational cost of HSE calculations, we 
perform the finite temperature calculations using the PBE functional
instead, and we expect that our results will be qualitatively correct.
Figure~\ref{fig:phasediagram} shows that, with increasing temperature
and for pressures above $0.5$~GPa, a band inversion occurs, inducing a
transition from a topological to a normal phase. Experimental
observation of a temperature-induced topological phase transition should
be simplest at pressures just above the zero-temperature transition
pressure. More generally, temperature-induced topological phase
transitions should be observable near critical values of other external
parameters driving a topological phase transition. The
pressure-temperature phase diagrams of Bi$_2$Te$_3$, Bi$_2$Se$_3$, and
Sb$_2$Te$_3$ are qualitatively similar to that of Sb$_2$Se$_3$. These
materials, however, are deep inside the topological part of the phase
diagram at ambient pressure so that unrealistically high temperatures 
would be needed to induce a transition to the normal state (see
Fig.~\ref{fig:tdep}).

Our results show that in the Bi$_2$Se$_3$ family of compounds,
increasing temperature favors the normal phase. This is at odds with
the theoretical prediction of
Refs.~\cite{elph_topological_prl,elph_topological_prb}, where it
was argued that higher temperature should favor topological phases. 
The focus there
is on electron-phonon coupling only, but we have shown that thermal
expansion is equally important. Even if only electron-phonon coupling
is considered, our calculations still show that temperature does not
favor topological phases in the Bi$_2$Se$_3$ family of compounds.
To understand the discrepancy, we note several possibilities. First, the 
analysis of Refs.~\cite{elph_topological_prl,elph_topological_prb} only includes
the so-called Fan term, which arises from treating the first-order
phonon-induced change in the Hamiltonian at second order in
perturbation theory. However, the Debye-Waller term, which arises from the 
second-order change in the Hamiltonian treated at first order in 
perturbation theory, has been shown to be equally important for the
calculation of the temperature dependence of band 
gaps~\cite{PhysRevLett.105.265501}. Our finite displacement approach
includes both terms. Second, the analysis of 
Refs.~\cite{elph_topological_prl,elph_topological_prb} assumes that
the two bands that participate in the band inversion are well separated 
in energy from the rest of the bands. This ensures that, for sufficiently
small gaps, the dominant contribution comes from intraband transitions,
which have a gap-opening tendency in topological insulators.
In the systems we have studied, the bands at $\Gamma$ involved in the
band inversion are not well separated in energy from other bands.
For example, the conduction band minimum of Bi$_2$Te$_3$ and Sb$_2$Te$_3$
is not even at $\Gamma$, where band inversion occurs. Therefore,
we do not expect that the analysis of 
Refs.~\cite{elph_topological_prl,elph_topological_prb} applies in our case. 
Nonetheless, there is no reason why temperature could not favor topological 
order in some other materials that obey the assumptions laid out in 
Refs.~\cite{elph_topological_prl,elph_topological_prb},
and it would certainly be interesting to identify some such examples.

% Conclusion
%\section{Summary}

In summary, we have shown that thermal expansion and electron-phonon
coupling drive the temperature dependence of the band structure of
topological insulators. Increasing temperature favors the normal state
in the Bi$_2$Se$_3$ family of compounds, and induces a topological phase
transition in pressurized Sb$_2$Se$_3$. Temperature-induced changes to
the band gaps reach $0.3$~eV at $600$~K, and are of similar size to the
changes induced by electron correlation in these materials. Open
questions remain, such as
the effects of thermally induced changes in electron occupation, and
the possible importance of non-adiabatic phenomena, especially
close to band inversions.

Our work shows that first-principles calculations of the effects of
temperature on topological materials are feasible, and that temperature
can have important effects on the band structure and topological order
of these materials. It should facilitate future work, for example in the
search for a room temperature Chern insulator.

\textit{Note added.} Antonius and Louie recently reported 
first-principles calculations of the temperature dependence of the
band structure of the topological insulator 
BiTl(S$_{1-\delta}$Se$_{\delta}$)$_2$~\cite{antonius_ti_temperature_arxiv}.

\acknowledgments

This work was funded by NSF grant DMR-1408838. B.M.~thanks Robinson
College, Cambridge, and the Cambridge Philosophical Society for a Henslow 
Research Fellowship.

% Mac
\bibliography{ti}

\end{document}